\documentclass[%
 reprint,
 amsmath,amssymb,
 aps,
]{revtex4-1}

\usepackage{lineno}
\usepackage{graphicx}
\usepackage{dcolumn}
\usepackage{bm}

\begin{document}


\title{Flavor-dependent EMC effect from a nucleon swelling model}

\author{Rong Wang$^{1}$}
\email{wangrong@ipno.in2p3.fr}
\author{Rapha\"el Dupr\'e$^{1}$}
\author{Yin Huang$^{2}$}
\author{Baiyang Zhang$^{3}$}
\author{Silvia Niccolai$^{1}$}

\affiliation{%
$^1$ Institut de Physique Nucl\'eaire, CNRS-IN2P3, Univ. Paris-Sud, Universit\'e Paris-Saclay, 91406 Orsay Cedex, France
}%

\affiliation{%
$^2$ School of Physics and Nuclear Energy Engineering, International Research Center for Nuclei and Particles in the Cosmos
and Beijing Key Laboratory of Advanced Nuclear Materials and Physics, Beihang University, Beijing 100191, China
}%

\affiliation{%
$^3$ Wigner Research Center for Physics, H-1121 Budapest, Hungary
}%


\date{\today}

\begin{abstract}
We present our results for a flavor-dependent EMC effect based on the nIMParton nuclear PDFs,
in which the $x$-dependence is described with a nucleon swelling model.
The nuclear correction from nucleon swelling is considered through
a modification of the initial valence quark distributions instead of a dynamical rescaling.
To probe the flavor-dependence of the model, the experimental observables
are calculated applying nIMParton nuclear modifications for various
experiments: parity-violating deep inelastic scattering on nuclear target,
pion-induced Drell-Yan, and W-boson production in proton-nucleus collisions.
In addition, we present the expected effect for the spectator-tagged deep inelastic
scattering process, which will be performed by the CLAS12 collaboration with the ALERT
detector.
\end{abstract}

\maketitle


\section{
\label{sec:intro}
Introduction
}

Since the discovery of the EMC effect \cite{emc-discovery}, both particle and
nuclear physics communities have struggled to understand the impact of
the nuclear medium and of the binding of nucleons on the nucleon structure.
The intricate connection between the perturbative and nonperturbative mechanisms
involved in these questions are typical of the underlying Quantum Chromodynamics (QCD) theory.
Nonetheless, numerous models have been proposed to understand the
EMC effect in the past decades (see the following reviews
\cite{review-Arneodo,review-Geesaman,review-Norton,review-Arrington,review-Hen,review-Malace}).
Generally, many models describe fairly well the main features of the
EMC effect. Therefore, new data about the EMC effect is crucial
to constrain the models and to understand the EMC effect.

Up to now, the $x$, $Q^2$, and $A$ dependence
of the EMC ratios have been investigated with a large number of experimental measurements.
Basically all the models depict well the $x$-dependence of the EMC ratios.
The $Q^2$-dependence of structure function ratio is found to be weak in experiments,
and the $Q^2$-dependence of nuclear parton distribution functions (PDFs) obeys
the QCD-based DGLAP evolution which governs the $Q^2$-dependence of free nucleon
PDFs as well \cite{Bickerstaff1986,Eskola-1998,Eskola-1999}. The recent measurement of the nuclear
dependence of the EMC effect at JLab \cite{Seely-2009} implies that the EMC effect
dominantly originates from the high virtuality or the high local density
\cite{Weinstein-2011,Hen-2012,review-Arrington,review-Hen}.
The detailed study of the flavor-dependence of the EMC effect is one of
the next directions for future experiments
\cite{FlavorDep_PVDIS,FlavorDep_DY,FlavorDep_WProd,review-Malace}.
Investigating the variations of the nuclear medium modifications for quarks of
different flavors opens a new window to test the various models.

The CBT model \cite{CBT-polarized,CBT-NuTeV,Bentz-2010,FlavorDep_PVDIS} is
the first model to bring out the nuclear effect difference between up quark
and down quark. In CBT model, the nuclear PDFs are determined using a confining
Nambu-Jona-Lasinio model, where the nucleon is approximated as a quark-diquark
bound state in the Faddeev equation \cite{CBT-polarized}. The nuclear effect
is implemented with the scalar and vector mean fields coupling to the quarks,
and the strength of the mean-fields are self-consistently determined using
an equation of state for nuclear matter. The isovector-vector mean field
$\rho^0$ arisen from neutron or proton excess in nuclei breaks down
$u_p(x)=d_n(x)$ and $d_p(x)=u_n(x)$ for bound nucleons, resulting in the
flavor-dependence of the EMC effect.

Recently, nIMParton (nuclear ``I'M Parton'') global analysis
studied the nuclear parton distributions with a nonperturbative input
which consists of only three valence quarks \cite{nIMParton-paper,nIMParton-webpage}.
Instead of adding degrees of freedom from nuclear physics,
the EMC effect in nIMParton analysis is produced from the
deformation of valence quark distributions at the input scale $Q_0^2$ due
to the mechanism of nucleon swelling. The influence of the nuclear interactions
or the mean-filed mesons are all reflected in the ``swelled" nucleon in the model.
According to the Heisenberg uncertainty principle, the larger confinement size
gives rise to smaller widths of the momentum distributions of partons.
The nuclear PDFs at high $Q^2$ are then dynamically generated from
the QCD-based evolution \cite{nIMParton-paper} with the modified valence quark distributions.
The nIMParton nuclear modification factors for up, down and strange
quarks present some differences. There are no initial strange quarks
at the input scale $Q_0^2$. All the strange quarks are generated from gluon splitting
in QCD evolution. This is why the nuclear medium modification of strange quark distribution
is different from that of valence quarks. The nuclear modifications of up and down
quarks manifest some difference, and it is due to the width difference between the up
valence quark distribution and the down valence quark distribution. Inside the proton,
the width of down valence quark distribution is narrower than that of up
valence quark distribution. With the same size of confinement increase,
the down valence quark distribution deforms more greatly to meet
the condition of Heisenberg uncertainty principle.

On the experimental side, several high energy scattering processes are suggested
to observe the flavor-dependence of the EMC effect \cite{FlavorDep_PVDIS,FlavorDep_DY,FlavorDep_WProd}.
They are the parity-violating deep
inelastic scattering (PVDIS) process with a polarized electron beam \cite{FlavorDep_PVDIS}, the pion-induced
Drell-Yan (DY) processes with pion beams \cite{FlavorDep_DY}, and the W-boson production with high energy
proton-nucleus collisions \cite{FlavorDep_WProd}. In these experiments, the sensitivities to the flavor-dependence
of the EMC effect are all discussed under the CBT model. It is worthwhile and important
to see also the predictions from other models.
In this paper, we show the predicted experimental observables of above experiments
using the flavor-dependent nuclear effect from nIMParton nuclear PDFs.
Although the flavor-dependence of nIMParton model is weak,
it provides a baseline to understand the nuclear isovector force in CBT model.

Another, recently proposed, method to access the flavor dependent structure functions
is to detect the low energy recoil nuclei. A program of such measurements
using A Low Energy Recoil Tracker (ALERT) combined
with the CLAS12 detectors has been approved at JLab
\cite{TaggedDVCS-inco,TaggedDVCS-cohe,TaggedEMC-proposal}.
With the spectator tagged, one knows the type of the nucleon struck by the high
energy probe. The EMC effect of the bound proton and the bound neutron
can then be independently measured, which could shed some lights on the isospin-dependence of the EMC effect.
In this work, we also give predictions for the EMC effect difference between the nuclear medium modified
proton and the nuclear medium modified neutron.

In Sec. \ref{sec:model}, we review the nucleon swelling model used to explain
the EMC effect. The size of nucleon swelling obtained from nIMParton analysis
is compared with the experimental measurements and the model calculations.
The nIMParton analysis based on the nucleon swelling and the flavor-dependence
of the nuclear effect are introduced in Sec. \ref{sec:nIMParton}. The experimental
observables of PVDIS process, pion-induced DY process and p-A collisions are shown
in Sec.  \ref{sec:PVDIS}, Sec. \ref{sec:DY} and Sec. \ref{sec:WProduction} respectively,
applying nIMParton nuclear PDFs. In Sec. \ref{sec:TaggedDIS}, we discuss
the potential of tagged-DIS to probe the flavor-dependence
of the EMC effect. Lastly, a brief summary is given in Sec. \ref{sec:summary}.

\section{
\label{sec:model}
Nucleon swelling and the EMC effect
}

The models of the EMC effect can be roughly classified in two categories:
conventional nuclear physics models and the QCD-inspired models \cite{review-Arneodo}.
The conventional nuclear models usually take into account the reduced
nucleon mass in medium or the virtuality, which gives the $x$-rescaling models
\cite{Canal1984,Staszel1984,Akulinichev1985,Frankfurt1987,Jung1988,Ciofi1999}
($x=Q^2/(2m_N\nu)$) and the off-shellness corrections \cite{Dunne1986,Gross1992,Kulagin1994,Kulagin2006,Kulagin2014}.
The QCD-inspired models usually require an increase of the quark confinement,
or a simple increase of nucleon radius (nucleon swelling).
A bigger nucleon equals a higher resolution power of the probe.
In the language of QCD evolution, the $Q^2$-rescaling \cite{PLB129.346,PLB134.449,Nachtmann1984,PRD31.1004,NPB296.582}
(an higher resolution power) is carried out to interpret the effect.

The nucleon swelling discussed in this work refers to the increase of the quark confinement size.
Such quark confinement enlargement is present in the multiquark cluster models
\cite{ClusterModel-1,ClusterModel-2,ClusterModel-3,ClusterModel-4,ClusterModel-5},
while a smaller quark deconfinement is predicted in
the Quark-Meson Coupling (QMC) model \cite{QMC-1,QMC-2,QMC-3},
and the nuclear potential model \cite{PRC35.1586,PLB329.164,JPG21.317}.
In the multiquark cluster model, the heavy nuclei favor the formations of large multi-nucleon
clusters containing $3N$ ($N=1,2,3...$) valence quarks.
In the QMC model, the size of the non-overlapping nucleon bag changes with the exchange
of the mean-field meson. In the potential model, the three-quark
quantum system are modified by the nuclear attractive potential.

There are a few experiments which indicated an increase
of the quark confinement radius in the nuclear medium \cite{PLB157.13,PRC.31.2184,PRL.60.2723,PRL.65.2110}.
The nucleon swelling is found to be small for the Helium-3 nucleus through
a quasi-elastic scattering experiment, which is smaller than 3-6 percentage \cite{PLB157.13}.
The other experiment, with kaon probe, hints to an increase of the
confinement up to 20\% in $^{12}$C and $^{40}$Ca \cite{PRC.31.2184}.
Furthermore, an interesting analysis of the data of hadron-nucleus interaction
shows that the effective cross section with bound nucleon is slightly larger than
that with free nucleon, which could imply a size increase as well \cite{ZPC.32.537}.

In nIMParton, the increase of the nucleon size is obtained from
a global analysis to the nuclear DIS data from worldwide facilities.
To reproduce the data, we find swellings of the nucleon radius of 0.8\%, 2\% and 8\% for
deuteron, $^3$He and $^{208}$Pb respectively \cite{nIMParton-paper}.
The size of the estimated nucleon swelling from experiments and from various models
are listed in Table \ref{tab:size_of_swelling}.
We see that several models predict such effects of a few percent,
such as the QMC model \cite{QMC-3}, the binding potential model \cite{PRC35.1586},
the Skyrmion model \cite{PLB329.164}, the quark-nucleon interaction model \cite{JPG21.317},
the chiral quark-soliton model \cite{PRL91.212301}, the chiral symmetry restoration model \cite{PLB176.469},
the weak stretching model \cite{ZPC.58.541}, the PLC-suppression model \cite{NPB250.143},
and the statistical model \cite{NPA828.390}.

\begin{table}[h]
\caption{The magnitudes of nucleon swelling inferred from experiments
and predicted from various models.}
\begin{center}
\begin{tabular}{ cc }
  \hline\hline
  experiment/model                                     &   size of nucleon swelling \\
  \hline
  quasielastic scattering \cite{PLB157.13}             &   $<3-6$\% for $^3$He\\
  K$^+$-nucleus scattering \cite{PRC.31.2184}          &   $10-30$\% for $^{12}$C and $^{40}$Ca\\
  nIMParton \cite{nIMParton-paper}                     &   $2.0-8.1$\% for $^3$He - $^{208}$Pb\\
  QMC \cite{QMC-3}                                     &   $5.5$\% for typical nuclei  \\
  binding potential \cite{PRC35.1586}                  &   a few \% for typical nuclei \\
  Skyrmion model \cite{PLB329.164}                     &   $3-4$\% \\
  quark-N interaction \cite{JPG21.317}                 &   $\sim2$\% for nuclear matter \\
  chiral quark-soliton \cite{PRL91.212301}             &   $\sim2.4$\% for heavy nuclei\\
  chiral symmetry \cite{PLB176.469}                    &   $<10$\% for nuclear matter\\
  N-N overlapping \cite{PLB134.449}                    &   $4.7-22$\% for $^3$He - $^{208}$Pb\\
  weak stretching \cite{ZPC.58.541}                    &   $4.5-9.4$\% for $^4$He - $^{208}$Pb\\
  PLC-suppression \cite{NPB250.143}                    &   $1-3$\% \\
  statistical model \cite{NPA828.390}                  &   $2.2-5.0$\% for $^4$He - $^{197}$Au\\
  quark-quark correlation \cite{NPA480.469}            &   $15$\% \\
  chrial quark-meson \cite{NPA510.689}                 &   $\sim19$\% for nuclear matter\\
  string model \cite{PLB178.285}                       &   $40$\% \\
  \hline\hline
\end{tabular}
\end{center}
\label{tab:size_of_swelling}
\end{table}

\begin{figure}[htp]
\centering
\includegraphics[width=0.48\textwidth]{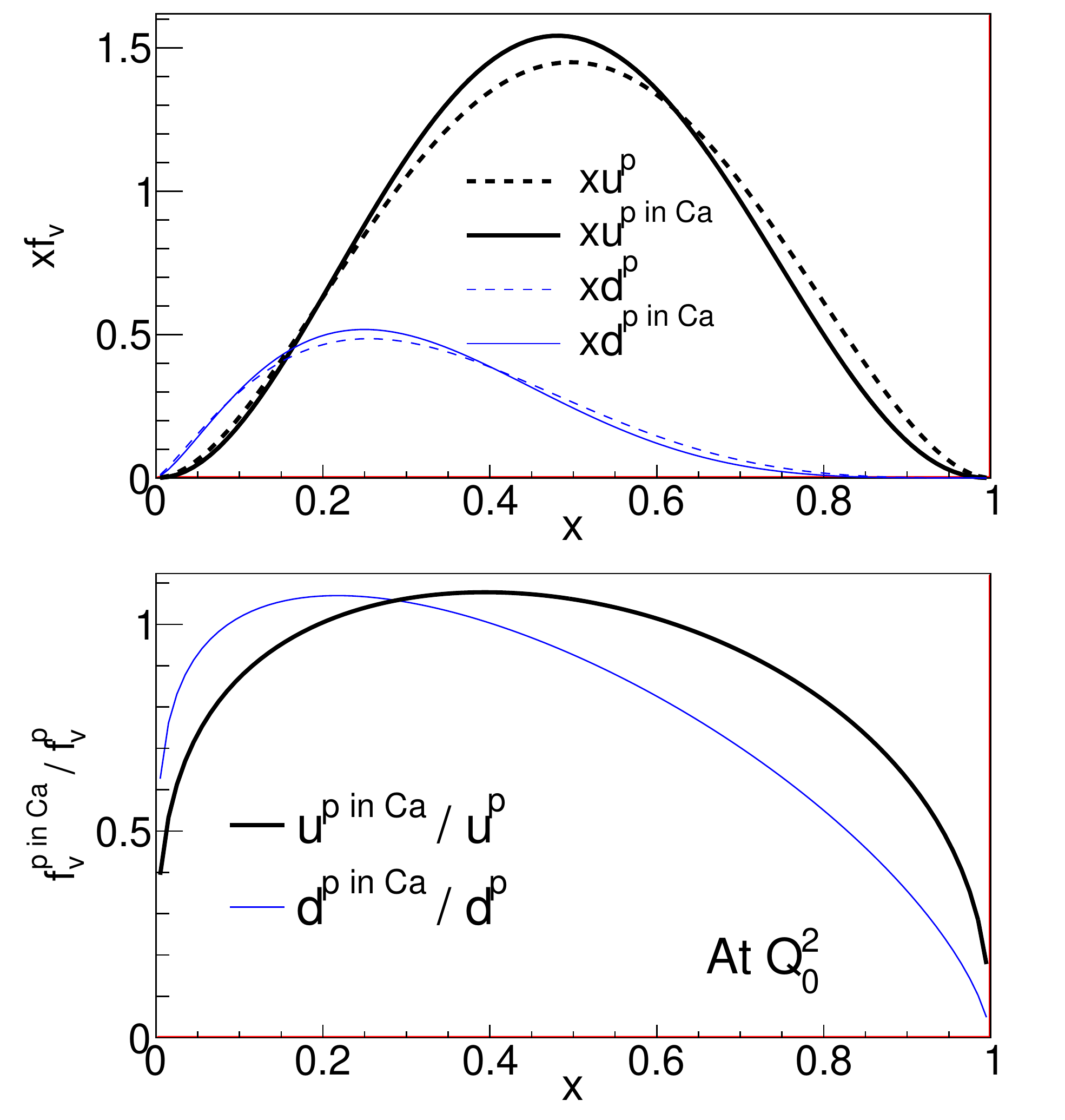}
\caption{
The valence quark distributions of free proton and the bound proton in $^{40}$Ca
at $Q_0^{2}=0.067$ GeV$^2$ are shown in the top panel. The EMC effect
from only the nucleon swelling is shown in the bottom panel.
}
\label{fig:CaInitialValence}
\end{figure}

Regardless of the origin of nucleon swelling, the valence quark distributions
are redistributed according to uncertainty principle to adapt for a larger
spacial uncertainty \cite{nIMParton-paper}.
In our model, all the medium modifications are reflected by this simple picture
of an increase of quark confinement, which changes the widths
of momentum distributions \cite{nIMParton-paper}.
The definition of the widths of valence distributions are written as,
\begin{equation}
\begin{aligned}
&\sigma(x_u)=\sqrt{<x_u^2>-<x_u>^2},\\
&\sigma(x_d)=\sqrt{<x_d^2>-<x_d>^2},\\
&<x_u>=\int_0^1 x\frac{u_v(x,Q_0^2)}{2}dx,\\
&<x_d>=\int_0^1 xd_v(x,Q_0^2)dx,\\
&<x_u^2>=\int_0^1 x^2\frac{u_v(x,Q_0^2)}{2}dx,\\
&<x_d^2>=\int_0^1 x^2d_v(x,Q_0^2)dx,
\end{aligned}
\label{WidthDef}
\end{equation}
and the deformation of the width of nuclear valence distribution is modeled
with a in-medium nucleon swelling parameter $\delta_A$. The momentum width
is inversely proportional to the nucleon size, as shown in the following equation,
\begin{equation}
\begin{aligned}
\frac{\sigma(x_q^A)}{\sigma(x_q^N)}=\frac{R_N}{R_{\text{in-medium}~N}}=\frac{1}{1+\delta_A},
(q=u,d).
\end{aligned}
\label{SwellingModel}
\end{equation}
For simplicity and because we lack information on the question,
the swellings of the bound proton and the bound neutron are identical in the model,
moreover the up and down valence quarks are confined in the same enlarged space.
In these conditions, the flavor-dependence of the EMC effect comes from the
difference in widths of the initial up and down valence quark distributions.
In consequence, when the confinement radius changes, the PDF ratio of down valence
quark is more affected than the ratio of up valence quark. This effect is mainly
affecting the range of $0.1<x<0.5$ at high $Q^2$ \cite{nIMParton-paper}.
Fig. \ref{fig:CaInitialValence}
shows the nuclear modifications on the initial valence quark distributions for $^{40}$Ca.
The EMC effect is the result by adjusting the widths of nuclear quark distributions.
Since the width of down valence quark distribution is narrower, the width of
$d^{\text{p~in~Ca}}/d^{\text{p}}$ curve is also narrower. In the calculation,
the valence quark distributions of both the free proton and the bound proton
are parameterized as $Ax^B(1-x)^C$, and they are required to satisfy the momentum sum rule
and the valence sum rule.

\section{
\label{sec:nIMParton}
nIMParton nuclear parton distribution functions
}

In IMParton global analysis, a nonperturbative input of only three valence quarks
is realized for free proton PDFs using DGLAP evolution with parton-parton
recombinations \cite{IMParton-paper,IMParton-webpage}. Based on the IMParton
analysis, the nIMParton analysis presents a global fit of nuclear PDFs with
the nucleon swelling assumption \cite{nIMParton-paper,nIMParton-webpage}.
Different from the traditional nuclear PDF analyses which use some arbitrary functions
to model the nuclear effect (the parton distribution ratios) with many parameters,
the nIMParton global fit is a model-dependent analysis with the nonperturbative input
consisting of only valence quarks, in order to better constrain the nuclear gluon distributions.
The nuclear gluon distributions are completely dynamically generated
in the DGLAP evolution with parton-parton recombinations, which are of small
bias theoretically.

\begin{figure}[htp]
\centering
\includegraphics[width=0.49\textwidth]{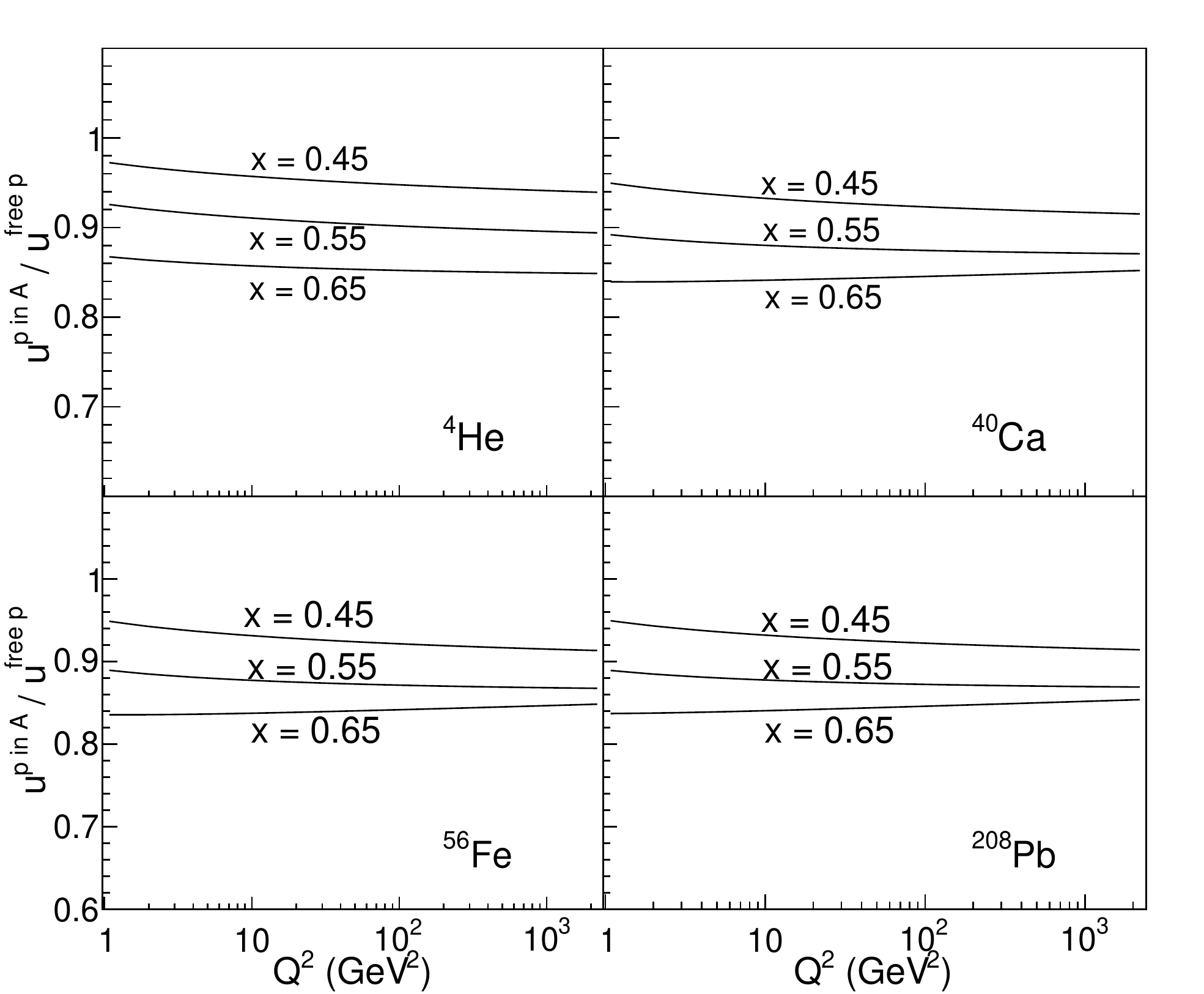}
\caption{
The $Q^2$-dependence of the nuclear effect for $^4$He, $^{40}$Ca, $^{56}$Fe and $^{208}$Pb.
}
\label{fig:nIMParton_Q2Dependence}
\end{figure}

Using much fewer parameters in the nIMParton analysis,
the nucleon swelling factor $\delta_A$ in Eq. (\ref{SwellingModel}) is modeled
to be proportional to the Residual Strong Interaction Enegy (RSIE),
as $\delta_A=\alpha\times RSIE/A$. $\alpha$ is a free parameter and fixed by the global fit.
$RSIE$ is simply the binding energy of strong interaction, defined as the nuclear binding
with the Coulomb part subtracted, of the formula $RSIE = B - B^{\text{Coul.}}$ \cite{WangPLB743.267}.
The nuclear binding $B$ is taken from the experimental measurement, and $B^{\text{Coul.}}$
is calculated with $-a_cZ(Z-1)A^{-1/3}$ ($a_c=0.71$ MeV). The $RSIE$ with different $A$, $Z$ and $N$
can easily be calculated. For nIMParton, the $A$, $Z$, and $N$ dependence
of the EMC effect are interpreted as the dependence on the binding energy of residual strong force (the $RSIE$).
The EMC effect of unmeasured nuclei also can be predicted, using $\alpha=0.00563$ MeV$^{-1}$
determined with nIMParton global fit. The nuclear PDFs at high $Q^2$ are calculated
using DGLAP equations with the modified nuclear valence distributions at $Q_0^2$.
The $Q^2$-dependence of the EMC effect is weak at high $Q^2$,
which is illustrated in Fig. \ref{fig:nIMParton_Q2Dependence}.

In addition to the nucleon swelling modeling above, a convolution formula
and the parton-parton recombination effect are taken to describe
the Fermi motion effect at large $x$ and the nuclear shadowing at small $x$,
respectively. No direct effects of virtual mean-field mesons are taken to
calculate the nuclear quark distributions in nIMParton analysis \cite{nIMParton-paper}.

\begin{figure}[htp]
\centering
\includegraphics[width=0.48\textwidth]{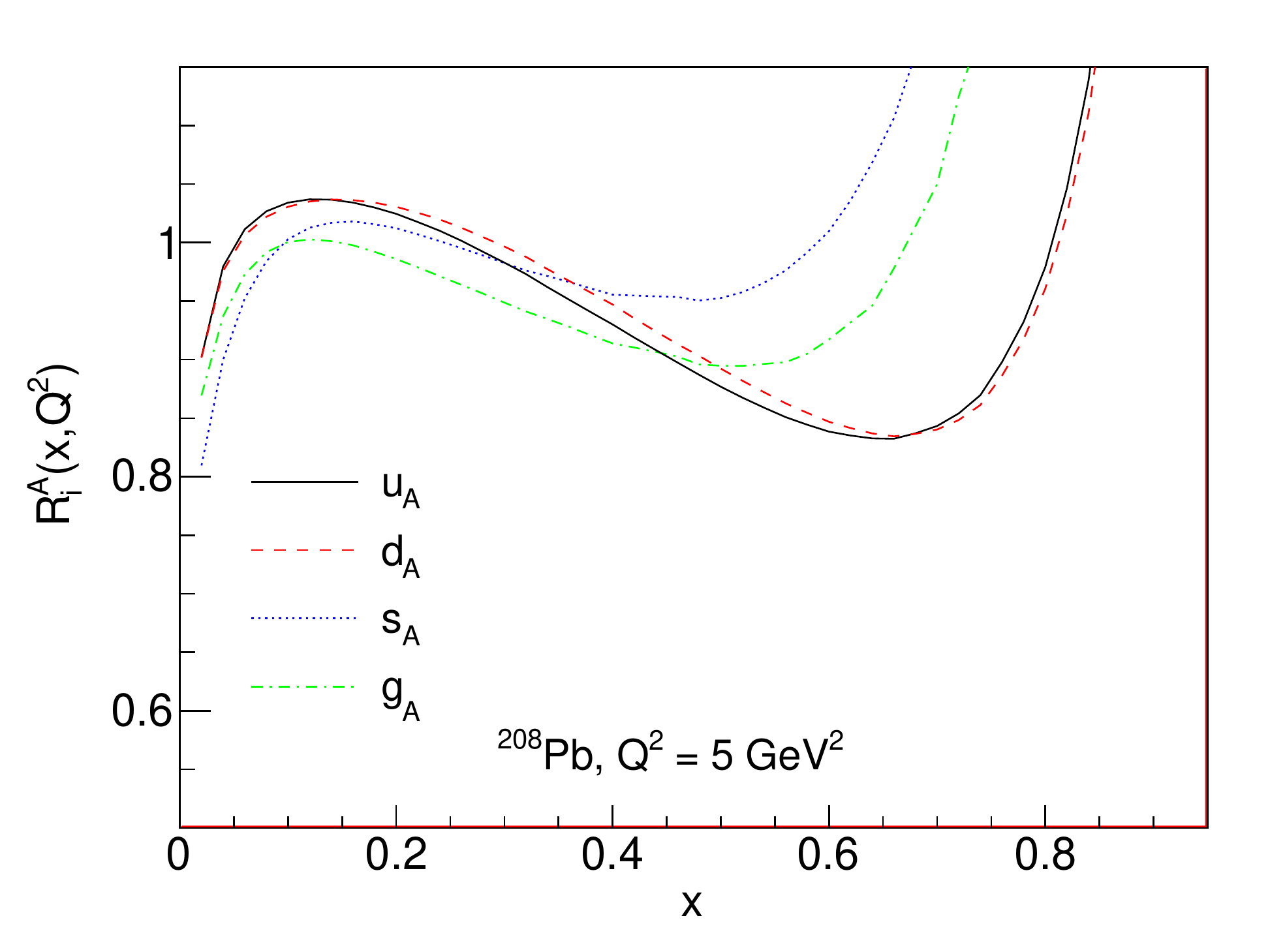}
\caption{Nuclear modification factors of different flavors
from nIMParton \cite{nIMParton-paper,nIMParton-webpage} are shown.
$R_i^A$ here is defined as $Af_i^A/[Zf_i^p+Nf_i^n]$,
in which $f_i^A$ is calculated with Eq. (\ref{eq:nPDFCalc}).
}
\label{fig:Ratios_nIMParton}
\end{figure}

In short, nIMParton is a model-dependent global fit of nuclear DIS data.
In the model, the EMC effect exhibit some difference between up valence quark
and down valence quark, which is due to the shape difference between up
valence distribution and down valence distribution. There are no sea quarks
and no gluons in the nonperturbative input at the initial scale $Q_0^2$ \cite{nIMParton-paper}.
All sea quarks and gluons are dynamically generated from the radiations of valence
quarks in the DGLAP evolution. Hence, it is not surprising that the EMC effect of
strange quark is different from valence quarks. In experiments, the nuclear
PDFs are the average distributions of all bound nucleons.
The nuclear PDF of flavor $i$ can be calculated
from the nIMParton nuclear modification factors with the following formula,
\begin{equation}
\begin{split}
f_i^A(x,Q^2)=\left[ZR_i^{bound~p}(x,Q^2)f_i^p(x,Q^2)\right.\\
\left.+(A-Z)R_i^{bound~n}(x,Q^2)f_i^n(x,Q^2)\right]/A,
\end{split}
\label{eq:nPDFCalc}
\end{equation}
where Z, A, $f_i^A(x,Q^2)$, $f_i^p$ and $f_i^n$ are atomic number, mass number,
nuclear PDF, proton PDF and neutron PDF respectively.
The nuclear modifications $R_i^{bound~p}(x,Q^2)$ and $R_i^{bound~n}(x,Q^2)$
can be accessed from the web \cite{nIMParton-webpage}.
For the calculations in this paper, the isospin symmetry is assumed between
proton and neutron, which implies $f_u^n = f_d^p$, $f_d^n = f_u^p$,
$f_{\bar{u}}^n = f_{\bar{d}}^p$ and $f_{\bar{d}}^n = f_{\bar{u}}^p$.

The results for the nuclear modification factors of Eq. (\ref{eq:nPDFCalc})
are shown in Fig. \ref{fig:Ratios_nIMParton}.
The differences among the ratios indicate the flavor-dependence of the
nuclear medium effect. Under the nIMParton data set, the differences are large
among the ratios of valence quark distribution, sea quark distribution, and gluon distribution
while the nuclear modification difference is small for up quark distribution
and down quark distribution in a nucleus, showing the maximum around $x=0.5$.
In the anti-shadowing region the sea quarks are slightly suppressed, which is consistent
with the Drell-Yan data from E772 \cite{E772-DY-data} and E866 \cite{E866-DY-data}.
These experiments measured the DY di-muon production in the range of $0.01<x<0.3$
for Carbon, Calcium, Iron and Tungsten nuclei, and found no enhancements of
the nuclear anti-quark distributions.

\section{
\label{sec:PVDIS}
Parity-Violating deep inelastic scattering
}

By using the polarized electron probe, the parity-violating DIS
experiment under high luminosity would present an important test on the difference
between the EMC effect of up quark and that of down quark \cite{FlavorDep_PVDIS}.
This idea to check the flavor-dependent modifications of nuclear medium
is to measure the difference between the traditional $F_2$ ratio and the $\gamma Z$
interference structure function ratio. The ratio definitions of the EMC effect for
both the traditional DIS and the $\gamma$Z interference structure functions are written as,
\begin{equation}
R^i=\frac{F_{2A}^i}{F_{2A}^{i,naive}}=\frac{F_{2A}^i}{ZF_{2p}^i+NF_{2n}^i},~(i=\gamma,~\gamma Z)
\label{eq:EMC_Ratios_Def}
\end{equation}
where $F_{2A}^{\gamma}$ and $F_{2A}^{\gamma Z}$ are the traditional unpolarized structure
function and the $\gamma Z$ interference structure function respectively.
The dominant term of the cross-section asymmetry between the positive and the negative
electron helicity is denoted as $a_2$ \cite{FlavorDep_PVDIS}, which is directly
connected to the ratio of $F_{2A}^{\gamma}$ and $F_{2A}^{\gamma Z}$.
Therefore the $F_{2A}^{\gamma Z}$ can be extracted combining the $a_2$ measurement
of PVDIS and the traditional $F_{2A}$ data.

\begin{figure}[htp]
\centering
\includegraphics[width=0.48\textwidth]{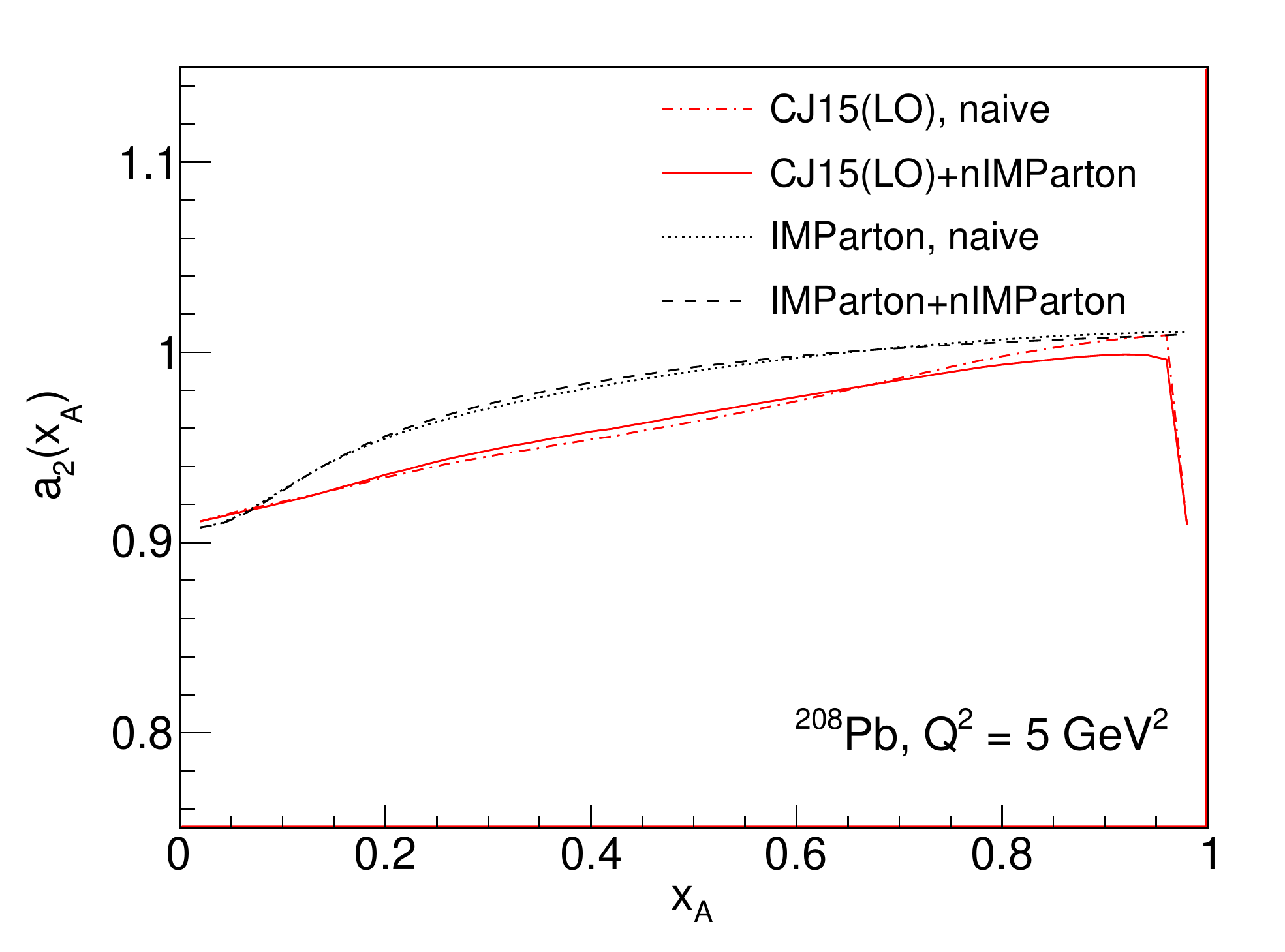}
\caption{The $a_2$ term of the analyzing power of longitudinally polarized
electron DIS scattering on $^{208}$Pb target. In the calculations,
the strange quark distribution and the heavy quark distributions are neglected.
CJ15(LO) PDF is taken from Refs. \cite{CJ15-paper,CJ15-webpage}.
IMParton PDF is taken from Refs. \cite{IMParton-paper,IMParton-webpage}.
nIMParton nuclear correction factor is from Refs. \cite{nIMParton-paper,nIMParton-webpage}.
}
\label{fig:a2_208Pb}
\end{figure}

\begin{figure}[htp]
\centering
\includegraphics[width=0.48\textwidth]{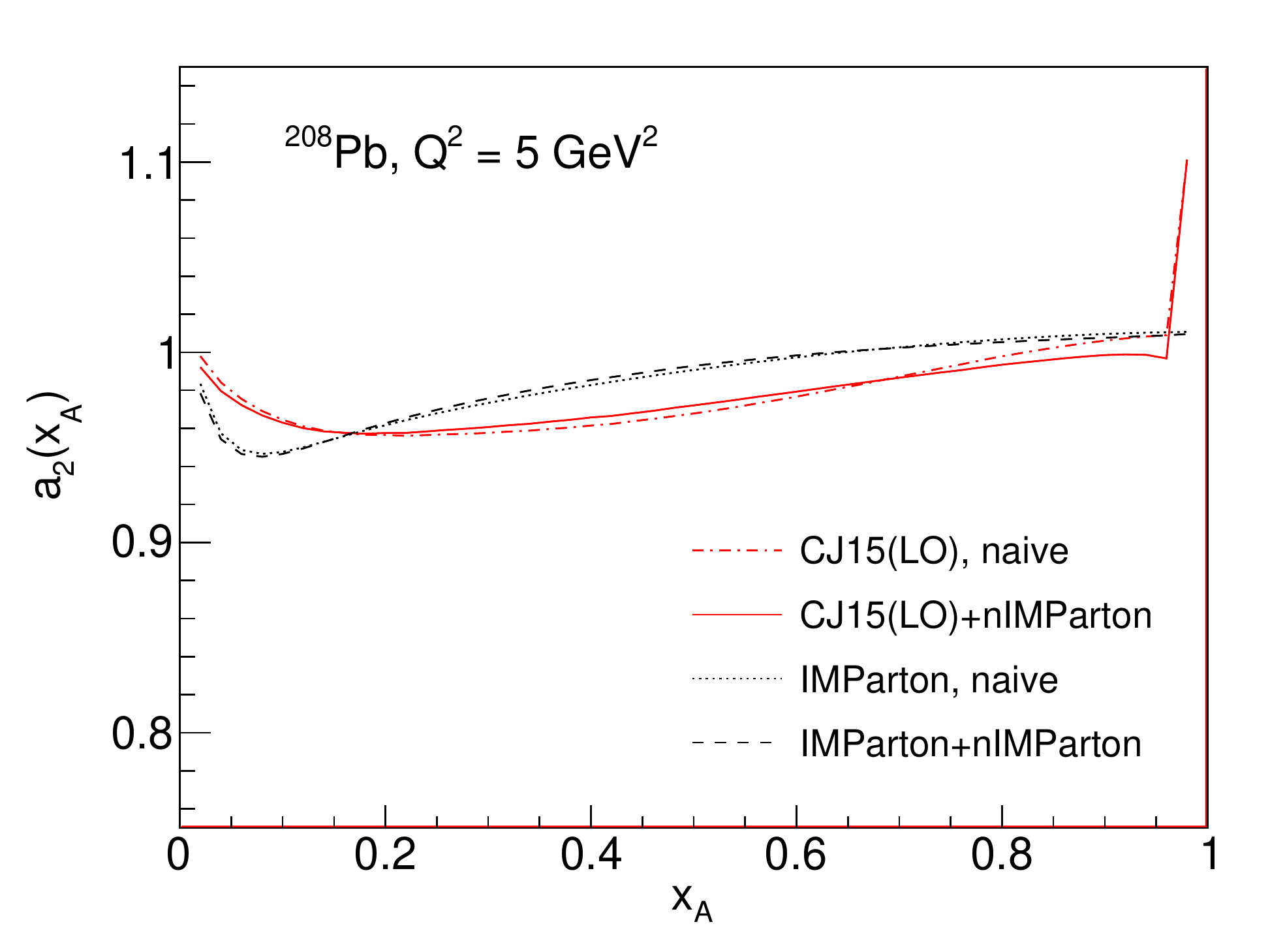}
\caption{The $a_2$ term of the analyzing power of longitudinally polarized
electron DIS scattering on $^{208}$Pb target. Up, down, and strange quark
distributions are all used in the calculations.
CJ15(LO) PDF is taken from Refs. \cite{CJ15-paper,CJ15-webpage}.
IMParton PDF is taken from Refs. \cite{IMParton-paper,IMParton-webpage}.
nIMParton nuclear correction factor is from Refs. \cite{nIMParton-paper,nIMParton-webpage}.
}
\label{fig:a2_208Pb_with_strange}
\end{figure}

Fig. \ref{fig:a2_208Pb} shows the $a_2$ of Lead using only up and down
quark distributions with the application of nIMParton nuclear modification
factors. The calculations of $a_2$ are given with the formula in Ref. \cite{FlavorDep_PVDIS},
which assumes $s+\bar{s}\ll u+d+\bar{u}+\bar{d}$. One can find that the $a_2$ value
actually depends on the PDF set used. Nevertheless, both PDF sets show small changes
of $a_2$ curves using nIMParton nuclear modifications, which is different from the
prediction of CBT model. There is an obvious difference between the naive $a_2$ and
the $a_2$ with CBT nuclear correction \cite{FlavorDep_PVDIS}. Moreover, the $x$-dependence of $a_2$
of $^{208}$Pb predicted from CBT model and that from nIMParton nuclear PDFs show clearly different behaviors.
The $a_2$ curve applying nIMParton nuclear PDFs is rather flat, while the $a_2$ curve goes up
quickly with $x$ approaching one in the CBT model (see Fig. 1 in Ref. \cite{FlavorDep_PVDIS}).
Fig. \ref{fig:a2_208Pb_with_strange} shows $a_2$ of Lead with strange quark
distribution included. Adding strange quark distribution changes much
$a_2$ in small $x$ region only. Therefore the $a_2$ measurement
in the valence region is feasible to distinguish the different models about
the flavor-dependent EMC effect.

\begin{figure}[htp]
\centering
\includegraphics[width=0.48\textwidth]{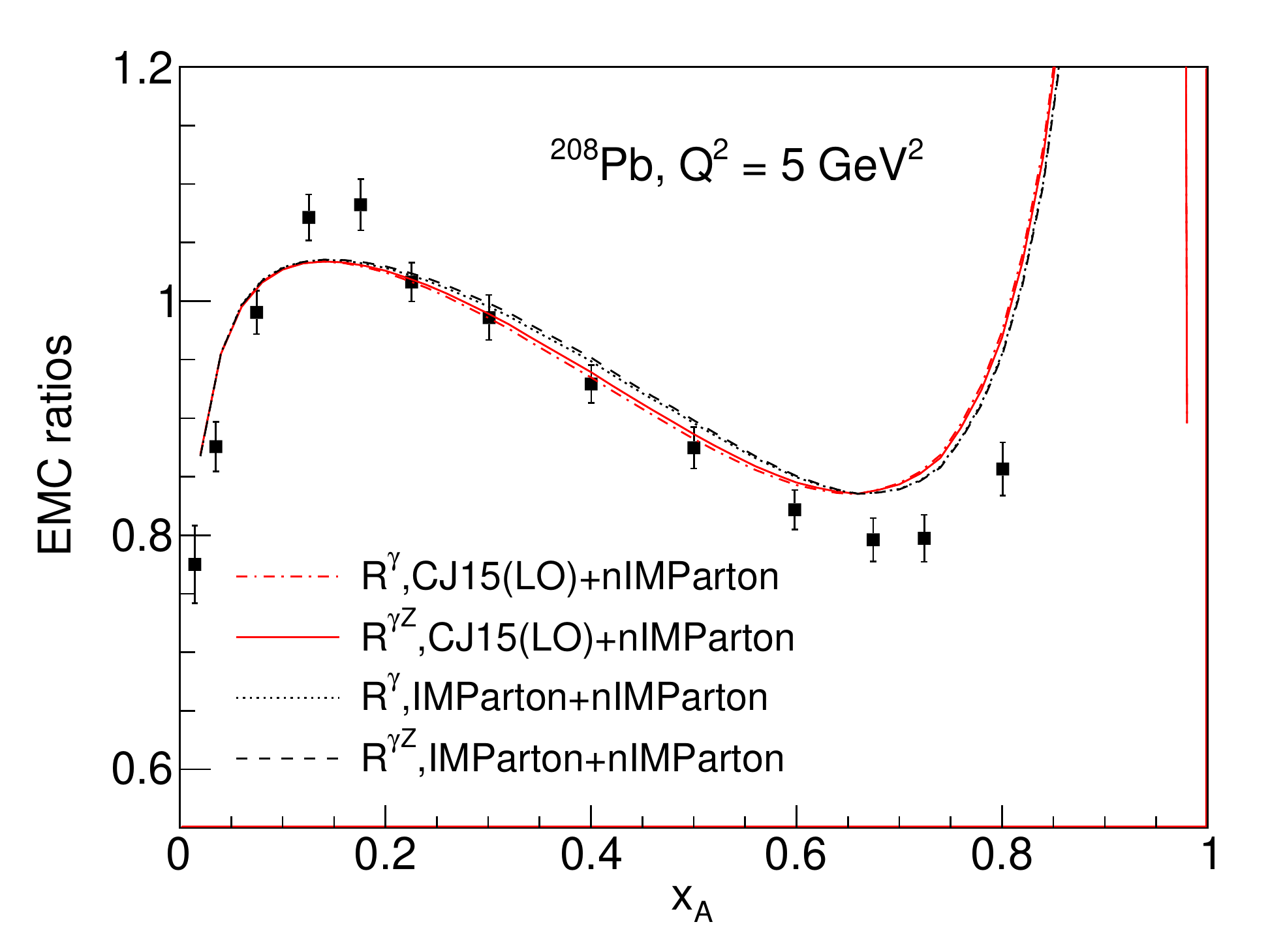}
\caption{The traditional DIS and the $\gamma$Z interference structure function
ratios of $^{208}$Pb to free nucleons. In the calculations, the strange quark
distribution and the heavy quark distributions are ignored.
CJ15(LO) PDF is taken from Refs. \cite{CJ15-paper,CJ15-webpage}.
IMParton PDF is taken from Refs. \cite{IMParton-paper,IMParton-webpage}.
nIMParton nuclear correction factor is from Refs. \cite{nIMParton-paper,nIMParton-webpage}.
The square points depict an extrapolation for infinite nuclear matter using a local
density approximation \cite{nuclearMatter_EMC}.
}
\label{fig:EMC_effect_208Pb}
\end{figure}

Fig. \ref{fig:EMC_effect_208Pb} shows the comparisons between the traditional
structure function ratio and the $\gamma$Z interference structure function ratio.
The formula to calculate these structure functions in terms of up and down
quark distributions can be found in Ref. \cite{FlavorDep_PVDIS}.
Based on nIMParton nuclear modifications, the difference
between $R^{\gamma}_{Lead}$ and $R^{\gamma Z}_{Lead}$ is trivial.
This conclusion is clearly different from that predicted by CBT model.
The CBT model predicts a noticeable difference between $R^{\gamma}_{Lead}$
and $R^{\gamma Z}_{Lead}$ based on the flavor-dependent nuclear force
(the $\rho_0$ mean field). The data points in Fig. \ref{fig:EMC_effect_208Pb}
show the extrapolated EMC ratios for infinite nuclear matter \cite{nuclearMatter_EMC}.
The heavy nucleus $^{208}$Pb can be viewed as the infinite nuclear matter approximately.
The predictions with nIMParton are consistent with the data.
The $R^{\gamma Z}_{Lead}$ extracted from PVDIS experiment is of significance
to check the predictions of the general nucleon swelling effect and
of the isovector field effect.

\section{
\label{sec:DY}
Pion-induced Drell-Yan process
}

Pion-induced Drell-Yan process is also a sensitive experimental tool
to probe the flavor-dependent EMC effect \cite{FlavorDep_DY}.
The DY cross section ratios which are sensitive to the
nuclear up and down quark distributions are denoted as,
\begin{equation}
\begin{split}
R_{\pm}^{DY}=\frac{\sigma^{DY}(\pi^+ + A)}{\sigma^{DY}(\pi^- + A)}
\approx \frac{d_A(x)}{4u_A(x)} \\
R_-^{DY,A/D}=\frac{\sigma^{DY}(\pi^- + A)}{\sigma^{DY}(\pi^- + D)}
\approx \frac{u_A(x)}{u_D(x)} \\
R_-^{DY,A/H}=\frac{\sigma^{DY}(\pi^- + A)}{\sigma^{DY}(\pi^- + H)}
\approx \frac{u_A(x)}{u_p(x)} \\
\end{split}
\label{eq:DY_Ratios_Def}
\end{equation}
where A, D, and H represent the nuclear, the deuteron and the hydrogen targets
respectively. $R_{\pm}$ measures the nuclear down quark to up quark ratio,
while $R_-$ measures the nuclear medium modification of up quark distribution.
The precise data of these DY cross-section ratios would provide
some stringent constrains to various models on the EMC effect.

\begin{figure}[htp]
\centering
\includegraphics[width=0.48\textwidth]{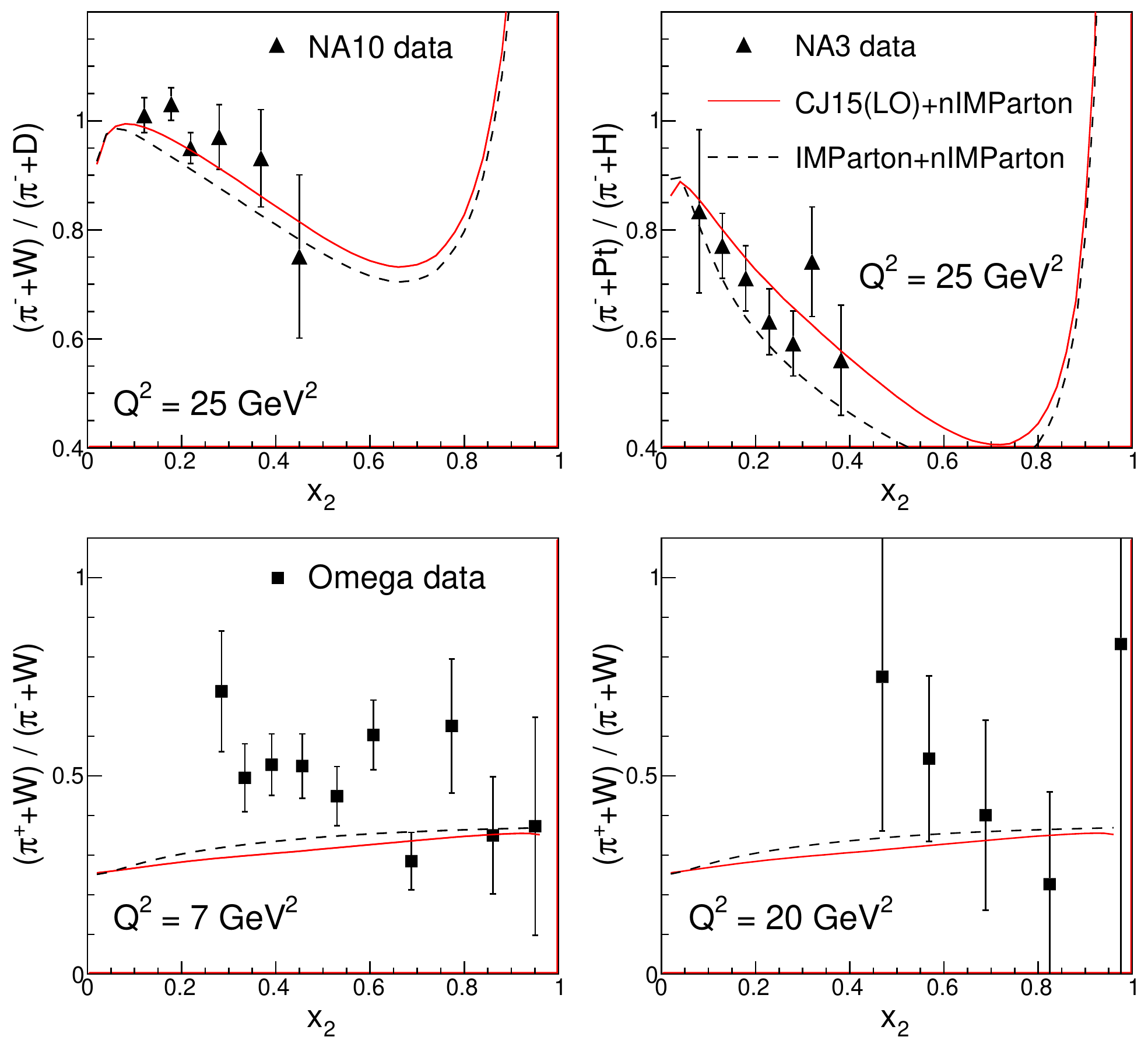}
\caption{The ratios between the cross-sections of
different pion-induced Drell-Yan processes.
CJ15(LO) PDF is taken from Refs. \cite{CJ15-paper,CJ15-webpage}.
IMParton PDF is taken from Refs. \cite{IMParton-paper,IMParton-webpage}.
nIMParton nuclear correction factor is from Refs. \cite{nIMParton-paper,nIMParton-webpage}.
The triangles represent the NA10 \cite{NA10-data} and NA3 \cite{NA3-data} data.
The squares represent the Omega data \cite{Omega-data}.
}
\label{fig:DrellYan_Ratios}
\end{figure}

The comparisons between the predictions from nIMParton nuclear modifications
and the existing pionic DY data are shown in Fig. \ref{fig:DrellYan_Ratios}.
The upper panels indicate that the nIMParton model
describe well the EMC effect of up quark distribution for both
Tungsten and Platinum targets. From the lower panels, we find that the nIMParton
nuclear modifications may not describe well the nuclear down quark to
up quark ratios. However the uncertainties of the $R_{\pm}$ data are quite big
up to date. The CBT model successfully interpret well all the data except
the NA10 data (see Fig. 2 in Ref. \cite{FlavorDep_DY}). We need the possible
future pion-induced Drell-Yan experiments to test the predictions and to
quantify the flavor-dependence of the EMC effect.

\section{
\label{sec:WProduction}
W-boson production in proton-nucleus collisions
}

Chang et al. suggest that it is another possible method to explore the flavor-dependent
EMC effect by measuring the differential cross-sections of W-boson production
in proton-nucleus collisions \cite{FlavorDep_WProd}. The experimental observables related to the topic
are the cross-section ratios which are defined as the followings:
\begin{equation}
\begin{split}
R^{+}_{A/D}(x_F)=\frac{\frac{d\sigma}{dx_F}(p+A\to W^+ + X)}{\frac{d\sigma}{dx_F}(p+D\to W^+ + X)}
\approx \frac{u_A(x_2)}{u_D(x_2)}, \\
R^{-}_{A/D}(x_F)=\frac{\frac{d\sigma}{dx_F}(p+A\to W^- + X)}{\frac{d\sigma}{dx_F}(p+D\to W^- + X)}
\approx \frac{d_A(x_2)}{d_D(x_2)}, \\
R^{\pm}_{A/D}(x_F)=\frac{\frac{d\sigma}{dx_F}(p+A\to W^+ + X)}{\frac{d\sigma}{dx_F}(p+A\to W^- + X)}
\approx \frac{\bar{d}_p(x_1)u_A(x_2)}{\bar{u}_p(x_1)d_A(x_2)}, \\
\end{split}
\label{eq:WProd_Ratios_Def}
\end{equation}
in which $x_F=x_1-x_2$ is the Feynman $x$ variable of the W-boson,
$x_1$ and $x_2$ are the momentum fractions carried by the partons
in the initial proton and in the initial nucleus respectively.
A and D denote the heavy nucleus and the deuteron respectively.

\begin{figure}[htp]
\centering
\includegraphics[width=0.48\textwidth]{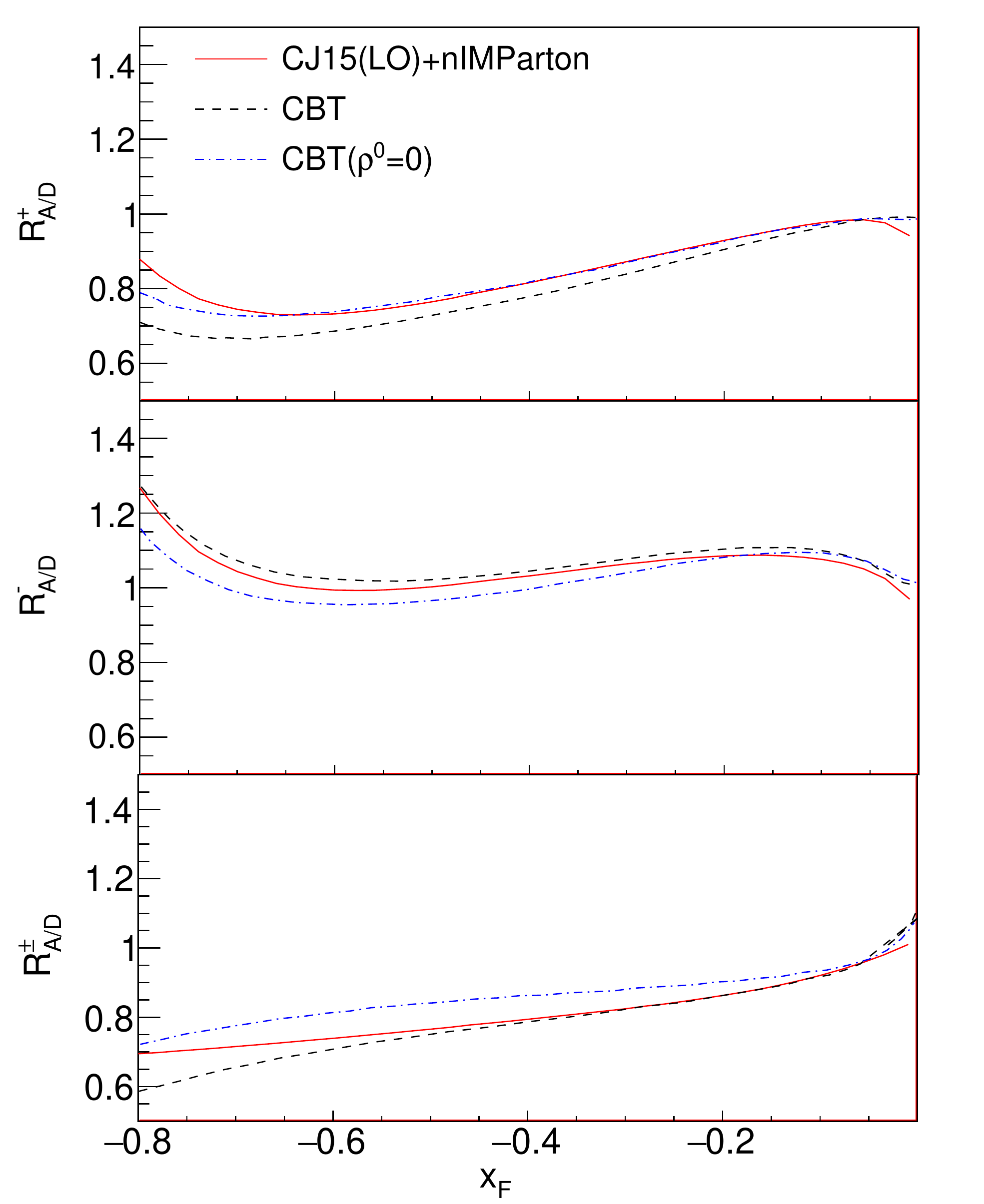}
\caption{nIMParton predictions of the cross-section ratios of W-boson
productions in proton-nucleus collisions (between $^{208}$Pb and deuteron, see text for explanations)
with the center-of-mass energy of $\sqrt{s}=5.520$ TeV,
compared with the CBT model \cite{CBT-polarized,CBT-NuTeV,Bentz-2010}.
CJ15(LO) PDF is taken from Refs. \cite{CJ15-paper,CJ15-webpage}.
nIMParton nuclear correction factor is from Refs. \cite{nIMParton-paper,nIMParton-webpage}.
}
\label{fig:W_Production_Ratios}
\end{figure}

Fig. \ref{fig:W_Production_Ratios} shows the predictions of the cross-section
ratios of W-boson production in different models for the proton-Lead collisions.
The $Q$ scale in the parton model calculations is chosen to be the W-boson mass scale.
It is interesting to find that the result based on nIMParton nuclear corrections
is between that of CBT model and that of CBT model without the
isovector meson force. These different model predictions can be verified
with the apparatuses at LHC or RHIC in the runs of high energy proton-nucleus collisions
under high luminosities.

\section{
\label{sec:TaggedDIS}
Spectator-tagged deep inelastic scattering
}

Spectator-tagged DIS from deuterium and $^4$He are proposed
to be measured using CLAS12 detectors combined with the ALERT detector
specialized in detecting the low energy spectator nuclei \cite{TaggedEMC-proposal}.
By tagging the nuclear recoil spectators ( $^4$He(e,~e'~$^3$H)X and $^4$He(e,~e'~$^3$He)X ),
we can probe the nuclear effect difference between the bound proton and the bound neutron.
The data of medium modified nucleons would provide some novel and important
tests on many models describing the EMC effect.

\begin{figure}[htp]
\centering
\includegraphics[width=0.48\textwidth]{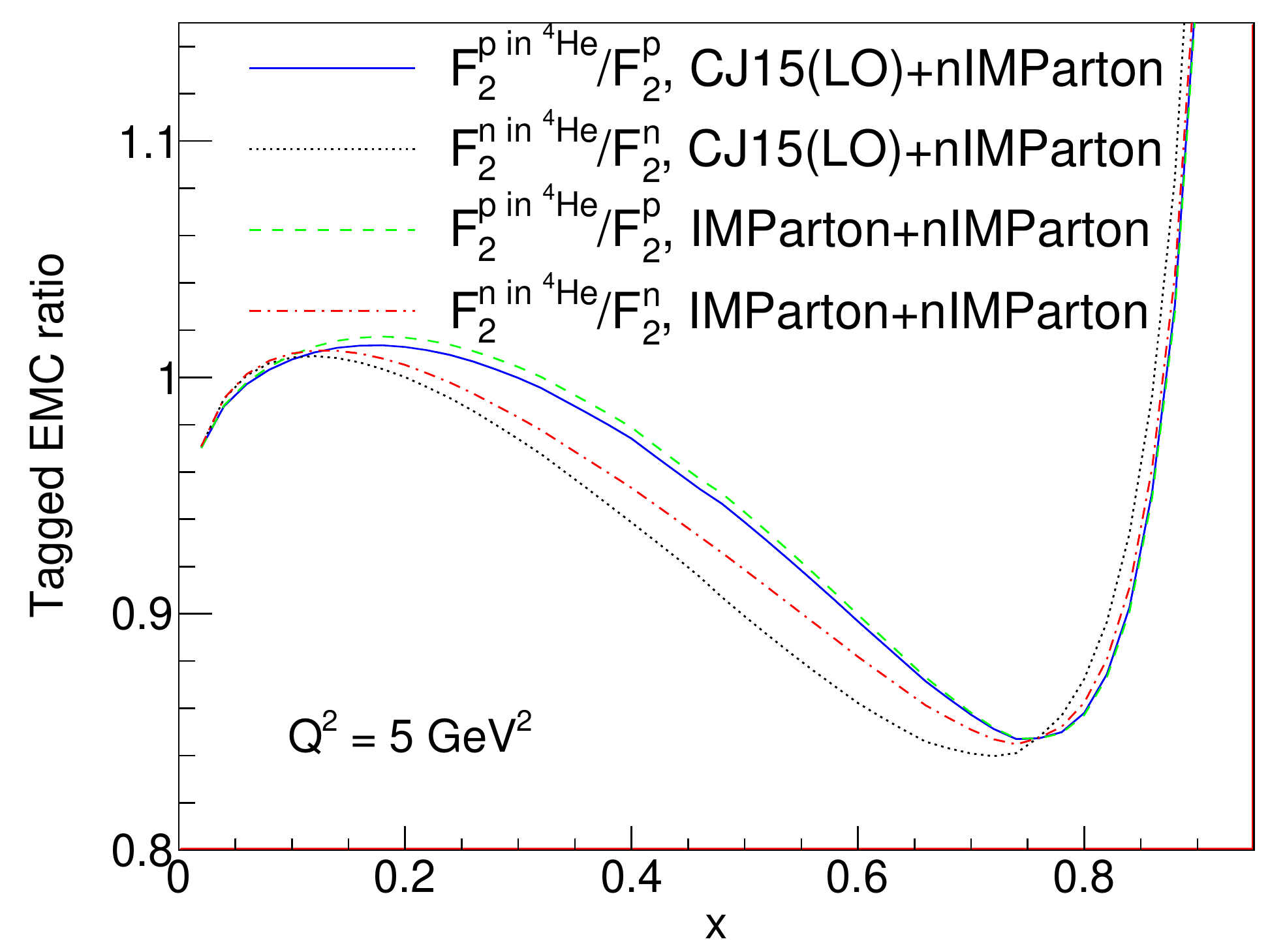}
\caption{Predictions of the spectator-tagged EMC ratios for $^4$He nucleus.
The PDFs of free proton and free neutron are taken from CJ15 \cite{CJ15-paper,CJ15-webpage}
and IMParton \cite{IMParton-paper,IMParton-webpage}.
The nuclear modifications of PDFs in bound nucleons are adopted from
nIMParton global analysis \cite{nIMParton-paper,nIMParton-webpage}.
}
\label{fig:Tagged_EMC_Ratios_of_He4}
\end{figure}

Measurement in DIS region displays the structure functions of nucleons.
In the leading order and ignoring the contributions of heavy quarks,
the proton structure function is expressed as
$F_2^p=x(\frac{4}{9}u^p + \frac{1}{9}d^p + \frac{1}{9}s^p)$.
Under the assumption of isospin symmetry, the neutron structure function is expressed as
$F_2^n=x(\frac{4}{9}u^n + \frac{1}{9}d^n + \frac{1}{9}s^n)=x(\frac{4}{9}d^p + \frac{1}{9}u^p + \frac{1}{9}s^p)$.
It is easy to see that the flavor-dependence of the nuclear modifications on quark distributions
could result in the difference of the EMC effect between medium modified proton
and medium modified neutron.
If the nuclear modifications on up and down quarks are the same,
then the EMC ratios for bound proton and bound neutron are identical
(ignoring the contribution of strange quark in the large $x$ region).

Fig. \ref{fig:Tagged_EMC_Ratios_of_He4} shows the EMC effects of bound proton
and bound neutron inside the $^4$He nucleus. In the calculations, the nuclear
modifications on parton distributions are taken from
nIMParton analysis \cite{nIMParton-paper,nIMParton-webpage}.
The EMC ratios for bound proton and for bound neutron exhibit some differences,
especially around $x=0.5$. Therefore, the tagged-DIS
experiment has the potential to test nIMParton predictions and the nucleon swelling
model used to soften the valence quark distributions.
To quantify the EMC effect of the bound neutron, the free neutron structure
function data is needed as the denominator. Fortunately, the state-of-the-art measurement of
the nearly free neutron structure function is currently realized by the BoNUS
Collaboration at JLab \cite{neutronPDF-1,neutronPDF-2,neutronPDF-3}.

\begin{figure}[htp]
\centering
\includegraphics[width=0.48\textwidth]{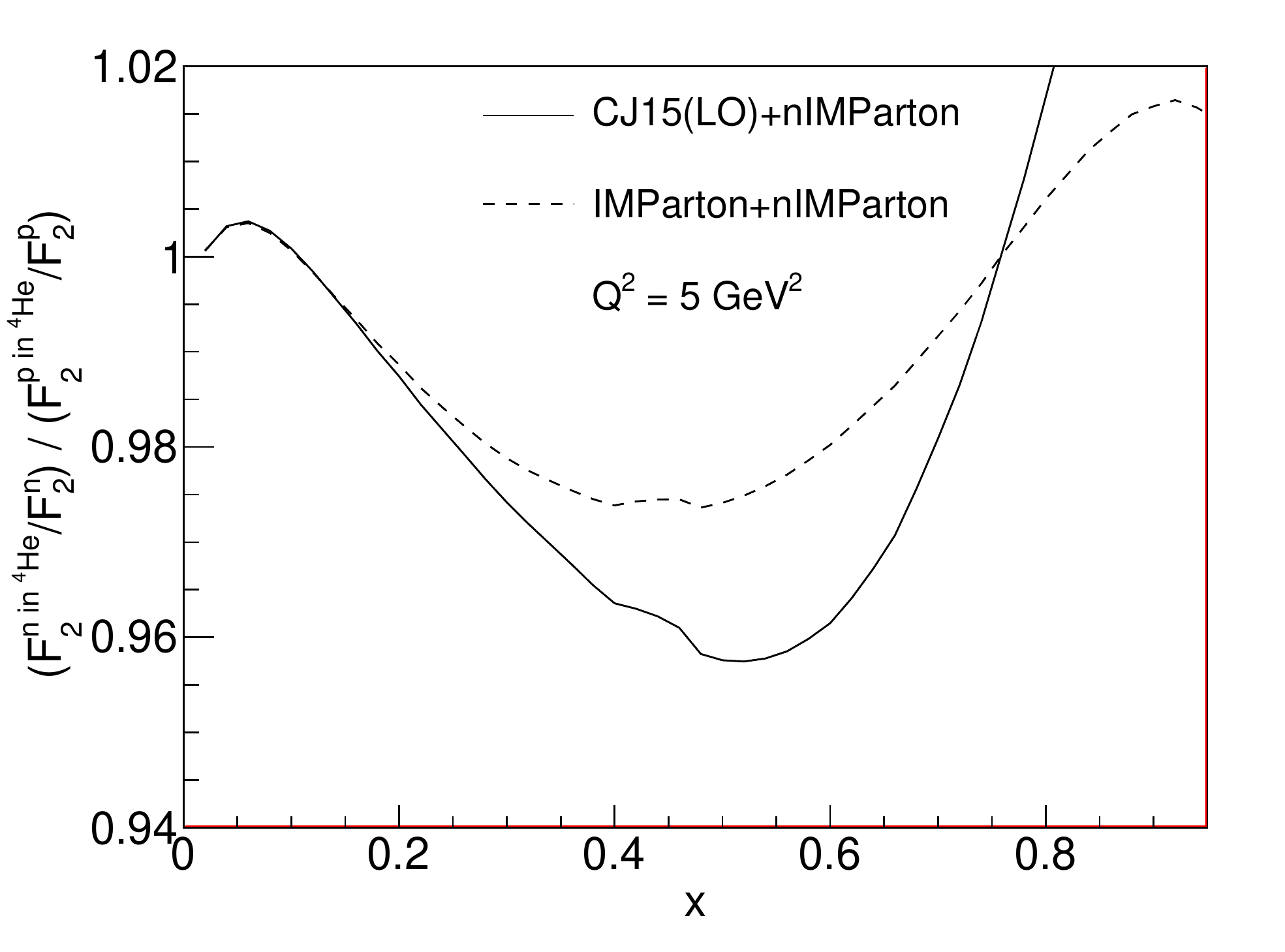}
\caption{The ratio of the nuclear modification factor on neutron $F_2$
to the nuclear modification factor on proton $F_2$, for $^4$He nucleus.
The PDFs of free proton and free neutron are taken from CJ15 \cite{CJ15-paper,CJ15-webpage}
and IMParton \cite{IMParton-paper,IMParton-webpage} for the calculations.
The nuclear modifications of PDFs in bound nucleons are adopted from
nIMParton global analysis \cite{nIMParton-paper,nIMParton-webpage}.
}
\label{fig:Nuclear_Modification_difference_p_and_n}
\end{figure}

To clearly demonstrate the magnitude difference of the EMC effect between
the bound proton and the bound neutron, the ratio of the nuclear EMC effect
is shown in Fig. \ref{fig:Nuclear_Modification_difference_p_and_n}.
It is shown that at $x$ around 0.5, the difference between the EMC ratios
of the bound proton and bound neutron is at the maximum of about 3\% relatively.
The ALERT experiment will be able to explore the variation
of the nuclear modification within the statistical error bars of 1 to 2\% \cite{TaggedEMC-proposal}.
Hence, the ALERT detector with CLAS12 would play an important role
in unveiling the issue on the isospin-dependent nuclear medium effect.

\section{
\label{sec:summary}
Summary
}

We discussed the new aspect of the EMC effect, the flavor-dependence,
using nIMParton nuclear modification factors. The forms of valence quark
distributions are modified according to the Heisenberg uncertainty principle \cite{nIMParton-paper}.
The nIMParton global analysis assumes the same enlargement of the
confinement size for both proton and neutron, and for both up valence quark
and down valence quark. Therefore, the isospin-dependence of the EMC effect
in the model is due to the fact that the down valence quark distribution
is narrower and softer than up valence quark distribution. The CBT model with the
isospin-dependent nuclear forces predicts larger nuclear effect difference
between up and down quark distributions than that of nIMParton,
for heavy nuclei with $N\neq Z$.

The flavor-dependent EMC effect based on nIMParton nuclear PDFs is
demonstrated with the predictions of various observables in the suggested
experiments of PVDIS, pion-induced Drell-Yan, W-boson production in p-A collisions,
and the tagged-DIS processes. The nIMParton predictions are consistent
with the Drell-Yan data decades ago. However we need more experiments
in the future to explicitly differentiate various models on EMC effect.
The experiment on CLAS12 with the ALERT detector at JLab is available to
test the models on the flavor dependence of the nuclear effect
in large $x$ region, which could provide a timely and critical
insight of the new aspect of the EMC effect.

One aim of the ALERT experiment is to measure the nuclear effect
of the mean-field nucleon and of the short-range correlated nucleon \cite{TaggedEMC-proposal}.
With the technique of tagging the recoil nuclei, the EMC effect as a function
of nucleon off-shellness can be deduced.
In this work, the EMC effect investigated is the average EMC effect of
the nucleons of different virtualities. If the confinement enlargement goes up as the local density
increase, the EMC effect should consequently enhanced for the case
of high momentum nuclear spectator recoil.

\end{document}